\documentclass{article}

\usepackage{amsmath}
\usepackage{PRIMEarxiv}
\usepackage{mathrsfs}
\usepackage{listings}
\usepackage{longtable}
\usepackage{tabularx}
\usepackage{forloop}
\usepackage{etoolbox}
\usepackage{multirow}
\usepackage{calc}
\usepackage{wrapfig,lipsum,booktabs} 
\usepackage{xtab} 
\usepackage[dvipsnames]{xcolor}

\usepackage[utf8]{inputenc} 
\usepackage[T1]{fontenc}    
\usepackage{hyperref}       
\usepackage{url}            
\usepackage{booktabs}       
\usepackage{amsfonts}       
\usepackage{nicefrac}       
\usepackage{microtype}      
\usepackage{lipsum}
\usepackage{fancyhdr}       
\usepackage{graphicx}       
\graphicspath{{media/}}     

\title{Comparing Different Transformer Model Structures for Stock Prediction}

\author{
  Qizhao Chen \\
  Graduate School of Information Science \\
  University of Hyogo \\
  Kobe, Japan\\
  \texttt{af24o008@guh.u-hyogo.ac.jp} \\
}

\begin{document}
\maketitle

\begin{abstract}
This paper compares different Transformer model architectures for stock index prediction. While many studies have shown that Transformers perform well in stock price forecasting, few have explored how different structural designs impact performance. Most existing works treat the Transformer as a black box, overlooking how specific architectural choices may affect predictive accuracy. However, understanding these differences is critical for developing more effective forecasting models. This study aims to identify which Transformer variant is most suitable for stock forecasting. This study evaluates five Transformer structures: (1) encoder-only Transformer, (2) decoder-only Transformer, (3) Vanilla Transformer (encoder + decoder), (4) Vanilla Transformer without embedding layers, and (5) Vanilla Transformer with ProbSparse attention. Results show that Transformer-based models generally outperform traditional approaches. Transformer with decoder only structure outperforms all other models in all scenarios. Transformer with ProbSparse attention has the worst performance in almost all cases. 
\end{abstract}


 \section{Introduction}

Stock price prediction is an important topic in finance. Many investors and researchers try to find good models to predict future stock movements. Traditional machine learning models like Support Vector Regression (SVR)~\cite{6572570} and Random Forest~\cite{9987903} have been used for this task. Later, deep learning models such as Long Short Term Memory (LSTM)~\cite{10392023} and Temporal Convolutional Network (TCN)~\cite{9851776} showed better results because they can better capture complex patterns in time series data.

Recently, Transformer models have become very popular. They were first designed for natural language processing tasks~\cite{vaswani2023attentionneed}, but now they are also used in time series forecasting~\cite{10825946}. Transformers use an attention mechanism, which helps the model focus on important parts of the input sequence. Many papers show that Transformers can outperform traditional models like LSTM in forecasting problems~\cite{mozaffari2024predictive,nguyen2023lightweight,9731073}. Because of this, more people are applying Transformers to stock price prediction.

However, most research uses only the basic Transformer model or a slightly modified version. Few studies directly compare different structures of Transformer models. It is still not clear how the choice of Transformer structure affects stock prediction performance. For example, some models only use the encoder~\cite{10825946}, while others use both encoder and decoder~\cite{Wang2023}. These differences can affect the results, but there is little research that looks at them carefully.

This paper studies five different Transformer structures for stock index prediction. The first model uses only the encoder part of the Transformer. The second model uses only the decoder part. The third model is the standard Transformer, which has both encoder and decoder. The fourth model is the standard Transformer without input embeddings. The fifth model is the standard Transformer but replaces the full attention with ProbSparse attention. This study also compares these models with LSTM, TCN, SVR, and Random Forest to see if Transformers really perform better.

This research selects the S\&P 500 index, one of the most widely followed equity indices representing the performance of 500 large-cap US companies, as the testing dataset. Daily stock closing prices were collected from May 1, 2015, to May 8, 2024. 

To make the prediction, this study uses sliding windows of different sizes, which are 5, 10, and 15 days. Each window is used to predict the stock index 1, 5, and 10 days into the future. Using different window sizes helps the models learn from different lengths of past information~\cite{756971}. Short windows (5 days) capture recent trends and quick market changes, while longer windows (10 and 15 days) provide a broader view of market behavior and smooth out short-term noise. This design allows us to check how the models perform over both short and long horizons, and whether they can adapt to different patterns in the time series.

In the previous work~\cite{10825946}, an ablation study of Informer model was conducted to predict an individual stock. In this study, the goal is to find out which Transformer structure works best for stock index prediction with different sliding window sizes. I believe this study can help other researchers and investors choose better model designs for stock forecasting tasks.

The remainder of this paper is structured as follows. Firstly, related work is listed (Section 2). The methodology is then described (Section 3). In Section 4, the experimental results are presented. Finally, Section 5 concludes the paper.

\section{Related Work}
Stock price prediction is a well-explored field in finance, with numerous models tested over time. Traditional methods, such as statistical models like linear regression~\cite{9943255} and ARIMA~\cite{7046047}, were among the first used for time series prediction, including stock prices. These methods are straightforward and easy to interpret but often fail to capture the complex, non-linear patterns typical of financial data.

The Transformer model, originally developed for natural language processing, has recently been adapted for time series forecasting, including stock price prediction. For example, Li et al.~\cite{li2023mastermarketguidedstocktransformer} propose a Market-Guided Transformer to predict stock prices. This model uses momentary and cross-time stock correlation as well as market information for automatic feature selection. Ji et al.~\cite{ji2024galformer} introduce a transformer-based model called Galformer with generative decoding and a hybrid loss function. The experimental results show superiority of Galformer over other traditional approaches. Xie et al.~\cite{electronics13214225} propose a Deep Convolutional Transformer model that combines convolutional neural networks, Transformers, and a multi-head attention mechanism.

Transformers are commonly employed in a full encoder-decoder setup for sequence-to-sequence tasks. However, for stock prediction, some researchers utilize only the encoder part of the Transformer, as the objective is typically to forecast the next value in a sequence rather than generate a completely new one. For example, Brugière and Turinici~\cite{10.1007/978-3-031-84460-7_33} use an encoder-only Transformer to forecast the S\&P 500 stock index. Lin and Tseng~\cite{electronics13112094} propose a Periodic Transformer Encoder (PTE) to improve the traffic time predictions. 

Variations of the Transformer, such as omitting the embedding layer to simplify the model or employing sparse attention mechanisms to reduce computational costs, have also been investigated. For example, Lu et al.~\cite{lu2023stockmarketindexprediction} use the Informer model to predict the stock index. Instead of using the full attention mechanism, the informer model utilizes a ProbSparse attention to reduce computation costs. Li et al.~\cite{li2020enhancinglocalitybreakingmemory} propose a LogSparse Transformer to reduce computational cost in time series forecasting. Haviv et al.~\cite{haviv2022transformerlanguagemodelspositional} show that casual Transformer Language Models without explicit positional encoding can still generate competitive results compared to standard models. Zuo et al.~\cite{zuo2024positioninformationemergescausal} also show that Transformer models with causal attention without positional embedding can still solve tasks that require positional information.

Despite increasing interest in Transformer models, few studies directly compare different Transformer architectures. It remains uncertain which structure is best suited for stock price prediction. Some studies propose that simpler architectures, like the encoder-only model~\cite{ogunfowora2023transformerbasedframeworkmultivariatetime}, may suffice for time series forecasting, while others contend that the full encoder-decoder Transformer offers superior results~\cite{cheng2024leveraging2dinformationlongterm}. Additionally, variations in attention mechanisms, such as ProbSparse attention, might influence model performance although it can reduce the computation cost.

Overall, while there is considerable interest in applying Transformer models to stock price prediction, there is a lack of understanding regarding how different Transformer structures and settings impact forecasting performance. This research aims to fill this gap by comparing five different Transformer variants and evaluating their performance across various sliding window sizes and prediction horizons.

\section{Methodology}\label{sec3}

\subsection{Data}
\begin{figure*}[!htbp]
    \centering
    \includegraphics[width=1\textwidth]{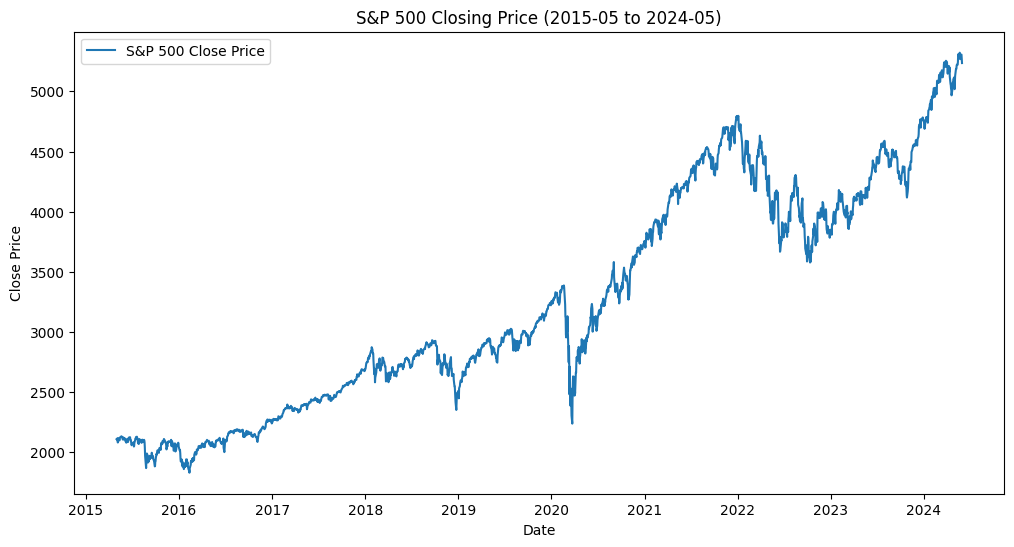}  
    \caption{S\&P 500 Close Price}
    \label{fig: sp500}
\end{figure*}

The dataset for this study includes daily closing prices for S\&P 500. The data covers the period from May 1, 2015, to May 8, 2024, and is gathered using the yfinance Python library. Table~\ref{tb: sp500 summary} shows the statistical summary of S\&P 500. Figure~\ref{fig: sp500} shows the time series of the S\&P 500 stock index from 2015 to 2024. The first 70\% of the data is used for training, while the remaining 30\% is reserved for testing. 

To train the models, a sliding window approach is used. This technique uses a fixed number of past days as input for predicting future stock prices. This study tests three different window sizes: 5, 10, and 15 days. For each window size, the model utilizes the previous 5, 10, or 15 days of stock data to forecast future prices. 

The prediction horizon is set to 1, 5, and 10 days ahead. This means that at each time step, the model is trained to predict the stock price for the next 1, 5, or 10 days based on the input sequence of historical stock prices.

\begin{longtable}{|l|l|}
\hline
\textbf{Statistic} & \textbf{Value} \\
\hline
\endfirsthead
\hline
\textbf{Statistic} & \textbf{Value} \\
\hline
\endhead
\hline
Ticker & \^GSPC \\
\hline
count  & 2286.000000 \\
\hline
mean   & 3251.590241 \\
\hline
std    & 932.150808 \\
\hline
min    & 1829.079956 \\
\hline
25\%   & 2447.845032 \\
\hline
50\%   & 2975.974976 \\
\hline
75\%   & 4124.402344 \\
\hline
max    & 5321.410156 \\
\hline
\caption{Statistical Summary of S\&P 500 Index}
\label{tb: sp500 summary}
\end{longtable}

\subsection{Model Setup}

Five different Transformer architectures are tested to evaluate their performance in stock prediction. The details of each model are described below.

\subsubsection{Transformer 1 (Encoder-only Transformer)}

As shown in Figure~\ref{fig: vanilla Transformer}, this model employs only the encoder component of the Transformer architecture. It processes the input sequence, such as historical stock prices, and generates a sequence representation used to predict future stock prices.

The encoder comprises a series of identical layers, each featuring two primary sub-layers: a multi-head self-attention mechanism and a position-wise fully connected feed-forward network. Each sub-layer includes residual connections followed by layer normalization to facilitate training and enhance convergence. Positional encodings are also added to the input embeddings to maintain information on the relative and absolute positions of tokens, allowing the model to capture sequential dependencies.

In this paper, the Encoder-only Transformer is configured with eight attention heads and three encoder layers.

\begin{figure*}[!htbp]
    \centering
    \includegraphics[width=1.1\textwidth]{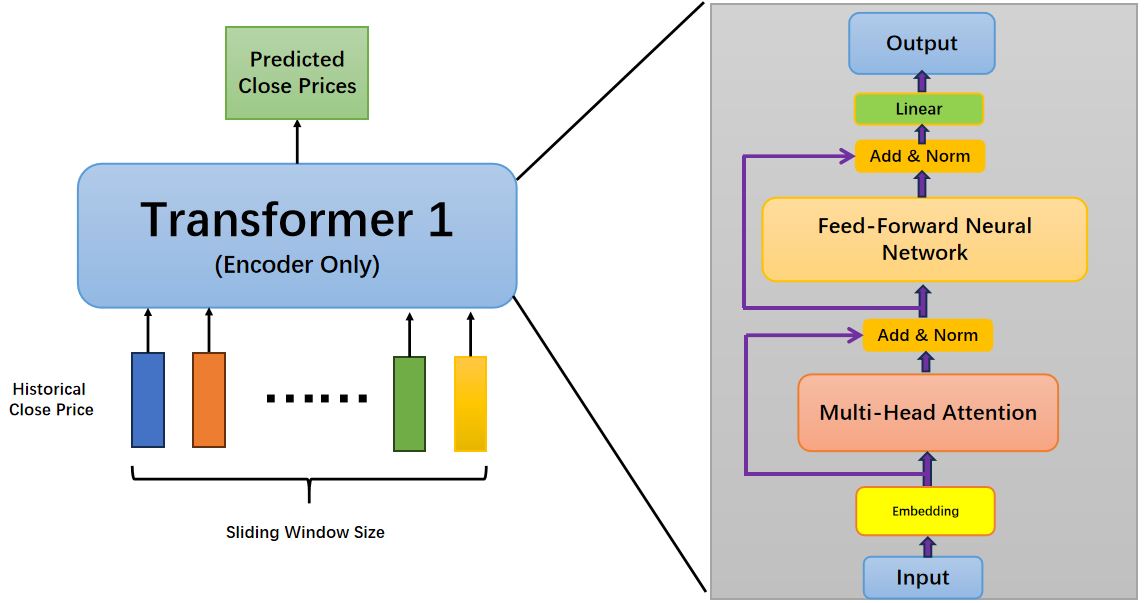}  
    \caption{Transformer 1 Structure}
    \label{fig:transformer 1}
\end{figure*}

\subsubsection{Transformer 2 (Decoder-only Transformer)}

\begin{figure*}[!htbp]
    \centering
    \includegraphics[width=1.1\textwidth]{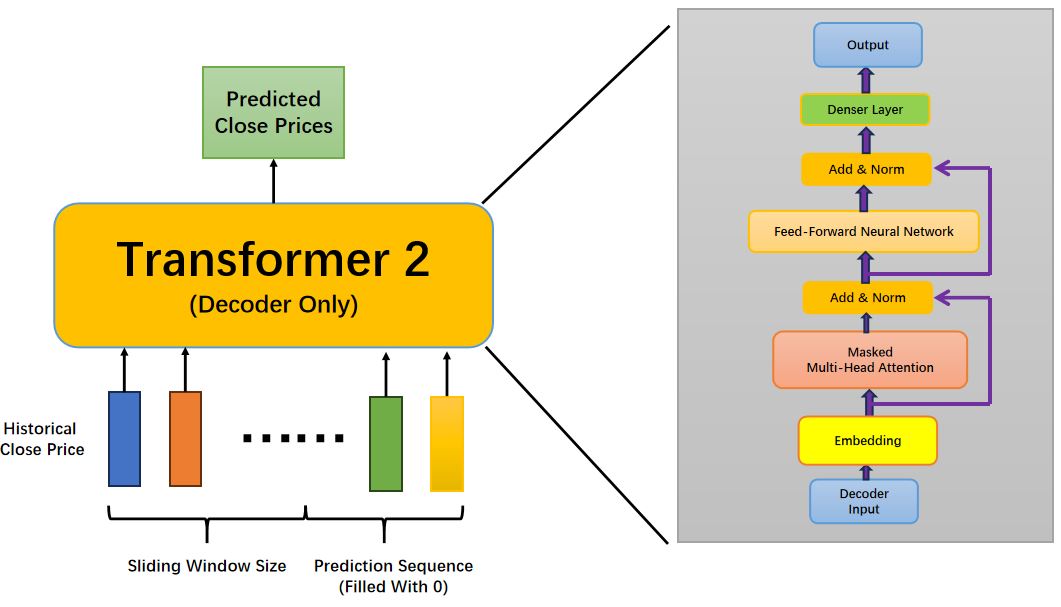}  
    \caption{Transformer 2 Structure}
    \label{fig:transformer 2}
\end{figure*}

As shown in Figure~\ref{fig:transformer 2}, this model uses only the decoder part of the Transformer architecture. The decoder consists of a series of identical layers, each featuring two main components: a masked multi-head self-attention mechanism and a position-wise fully connected feed-forward network. The masked self-attention ensures that each position in the decoder can only focus on previous positions in the output sequence, maintaining the autoregressive property during both training and inference. Residual connections and layer normalization follow each sub-layer to enhance stable and efficient training. Positional encodings are also added to the input embeddings to convey information about the position of tokens within the sequence.

In this paper, the Decoder-only Transformer is configured with eight attention heads and two decoder layers.

\subsubsection{Transformer 3 (Vanilla Transformer)}
\begin{figure*}[!htbp]
    \centering
    \includegraphics[width=1.1\textwidth]{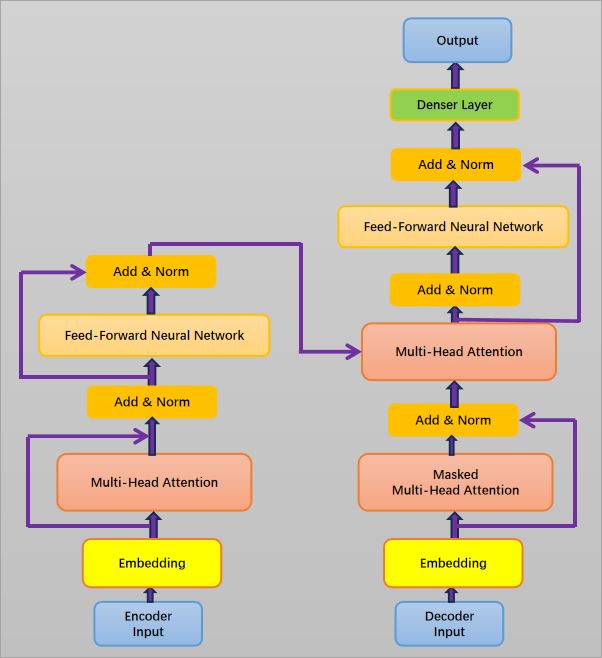}  
    \caption{Vanilla Transformer}
    \label{fig: vanilla Transformer}
\end{figure*}

As is shown in Figure~\ref{fig: vanilla Transformer}, this is the standard Transformer model that uses both the encoder and decoder. The encoder processes the input sequence, and the decoder generates the prediction for the future stock price.

In the Vanilla Transformer architecture, each encoder layer consists of two main components: a multi-head self-attention mechanism and a feed-forward neural network. Both components are followed by residual connections and layer normalization, which help preserve the original input's integrity and stabilize training. Decoder layers have a similar structure but include an extra encoder-decoder attention mechanism. This allows the decoder to focus on relevant parts of the encoder's output. To incorporate information about the position of tokens within the sequence, positional encodings are added to the input embeddings, enabling the model to effectively capture the sequence order.

In this paper, the Vanilla Transformer model is configured with eight attention heads, three encoder layers, and two decoder layers.

\subsubsection{Transformer 4 (Vanilla Transformer without Embedding)}

This model is similar to Transformer 3 but removes the embedding layer. The input features are used directly without transforming them into embeddings, which simplifies the model architecture.

\subsubsection{Transformer 5 (Vanilla Transformer with Probsparse Attention)}

\begin{figure*}[!htbp]
    \centering
    \includegraphics[width=0.7\textwidth]{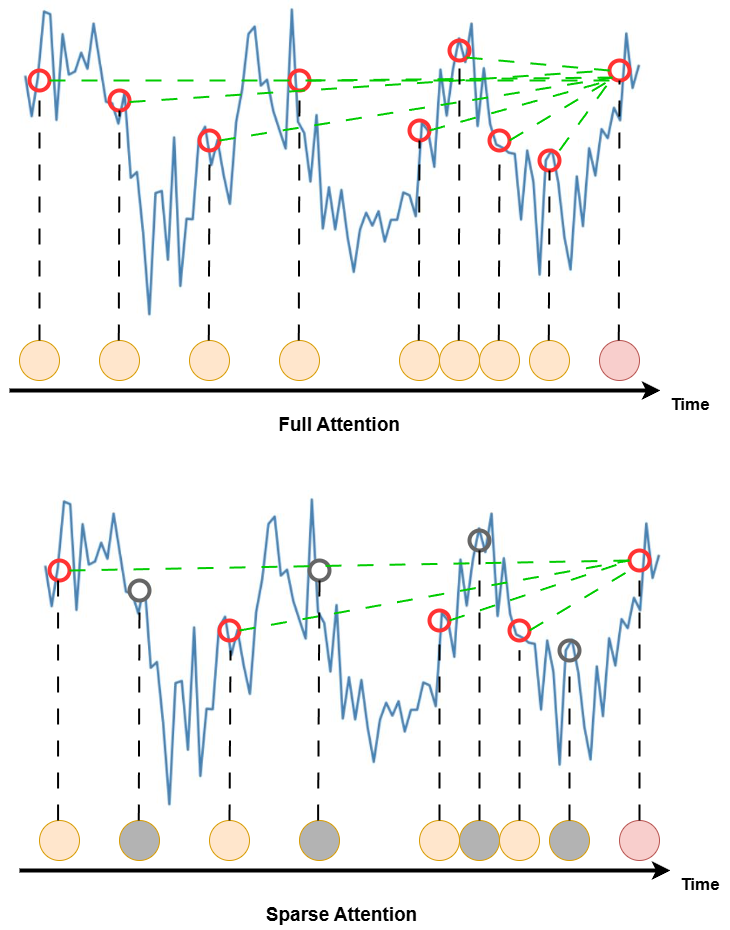}  
    \caption{Full Attention and Sparse Attention}
    \label{fig: sparse attention}
\end{figure*}

This model uses a vanilla Transformer with a sparse attention mechanism. In standard full attention, each token attends to every other token in the sequence, leading to quadratic computational complexity with respect to the sequence length. In contrast, as shown in Figure~\ref{fig: sparse attention}, ProbSparse attention focuses only on the most relevant parts of the input sequence. This selective attention reduces computational costs and makes the model more efficient without significantly sacrificing performance.

Specifically, the ProbSparse attention mechanism introduces three key components that work together.

First, the sparsity measure in Equation~\eqref{eq:sparsity_measure} evaluates how concentrated the attention distribution is for each query. It calculates the difference between the log-sum-exp of all attention scores and their average value. Queries showing larger differences tend to have more peaked distributions, focusing on just a few key elements.

However, computing this exact measure for every query would be computationally expensive. That's where Equation~\eqref{eq:approx_sparsity_measure} comes into play. This practical approximation uses only two simple statistics from a small subset of keys: the maximum attention score and the average score. Despite its simplicity, it reliably identifies which queries have sparse attention patterns.

Finally, the sparse attention operation itself appears in Equation~\eqref{eq:prob_sparse_attention}. It applies full attention computation only to selected queries, typically those with the top-$u$ sparsity scores. For all other queries, the mechanism takes a shortcut by simply passing through the corresponding value unchanged. This selective approach significantly reduces computation while maintaining the model's ability to focus on important relationships.

\begin{align}
M_{\text{sparse}}(q_i, K) &= \ln\sum_{j=1}^L e^{q_i k_j^\top/\sqrt{d}} - \frac{1}{L}\sum_{j=1}^L \frac{q_i k_j^\top}{\sqrt{d}} \label{eq:sparsity_measure} \\
\bar{M}_{\text{sparse}}(q_i, \tilde{K}) &= \max_j \left\{ \frac{q_i k_j^\top}{\sqrt{d}} \right\} - \frac{1}{\tilde{L}}\sum_{j=1}^{\tilde{L}} \frac{q_i k_j^\top}{\sqrt{d}} \label{eq:approx_sparsity_measure} \\
\text{SA}_{\text{sparse}}(q_i, K, V) &=
\begin{cases} 
\sum_{j=1}^L \frac{e^{q_i k_j^\top/\sqrt{d}}}{\sum_l e^{q_i k_l^\top/\sqrt{d}}}v_j, & \text{if } \bar{M}_{\text{sparse}}(q_i, \tilde{K}) \in \text{Top-}u \\
v_i, & \text{otherwise}
\end{cases} \label{eq:prob_sparse_attention}
\end{align}

\noindent where:
\begin{itemize}
\item For Equation~\eqref{eq:sparsity_measure}:
  \begin{itemize}
  \item $q_i$: The $i$-th query vector
  \item $k_j$: The $j$-th key vector
  \item $d$: Dimension of key/query vectors
  \item $L$: Total number of key vectors
  \item $\ln\sum e^{q_i k_j^\top/\sqrt{d}}$: Log-sum-exp of attention scores
  \item $\frac{1}{L}\sum \frac{q_i k_j^\top}{\sqrt{d}}$: Average attention score
  \end{itemize}

\item For Equation~\eqref{eq:approx_sparsity_measure}:
  \begin{itemize}
  \item $\tilde{K}$: Sampled subset of keys ($\tilde{K} \subset K$)
  \item $\tilde{L}$: Number of sampled keys ($\tilde{L} \ll L$)
  \item $\max_j\{\cdot\}$: Maximum attention score in the subset
  \end{itemize}

\item For Equation~\eqref{eq:prob_sparse_attention}:
  \begin{itemize}
  \item $v_j$: The $j$-th value vector
  \item $\text{Top-}u$: Top $u$ queries with highest sparsity scores
  \item $\frac{e^{q_i k_j^\top/\sqrt{d}}}{\sum_l e^{q_i k_l^\top/\sqrt{d}}}$: Softmax attention weights
  \end{itemize}
\end{itemize}

\subsection{Baseline Models}

To compare the performance of different Transformer architectures, four traditional models are used as benchmarks: LSTM, TCN, SVR, and Random Forest (RF). Each model is briefly described below.

\subsubsection{Long Short Term Memory (LSTM)}
LSTM is a type of recurrent neural network (RNN) designed to learn long-term dependencies in sequential data. It employs special units known as memory cells capable of retaining information over extended periods. Each memory cell contains input, output, and forget gates to regulate the flow of information. LSTM networks are extensively used in time series forecasting due to their ability to capture both short-term and long-term trends effectively. 

\subsubsection{Temporal Convolutional Network (TCN)}
TCN is a convolutional neural network designed for sequence modeling tasks. Unlike traditional convolutional networks, TCNs employ causal convolutions, ensuring that predictions at any time step depend solely on previous steps, not future ones. Additionally, TCNs utilize dilated convolutions, enabling a large receptive field while maintaining a manageable number of parameters. These networks have demonstrated strong performance in time series forecasting and offer benefits in training speed and stability over RNN-based models.

\subsubsection{Support Vector Regression (SVR)}
SVR is a supervised learning algorithm derived from Support Vector Machine (SVM) techniques. It seeks to identify a function that deviates from the actual observed targets by no more than a specified margin, while remaining as flat as possible. SVR is renowned for handling high-dimensional data and modeling non-linear relationships through kernel functions. In this study, I employ SVR with a radial basis function (RBF) kernel to forecast stock prices. As SVR does not inherently manage sequential data, the input is transformed into feature vectors using the sliding window method.

\subsubsection{Random Forest (RF)}
Random Forest is an ensemble learning technique that constructs multiple decision trees during training and provides the average prediction from these trees. This approach enhances predictive accuracy and mitigates overfitting through bagging (bootstrap aggregating) and random feature selection at each node split. Random Forests are resilient to noise and capable of modeling intricate non-linear relationships.

\subsection{Evaluation Metrics}

To evaluate the performance of the models, this study uses the following metrics.

\subsubsection{Mean Absolute Error (MAE)}
Mean Absolute Error measures the average magnitude of the errors in a set of predictions, without considering their direction. It is calculated as:
\begin{equation}
\text{MAE} = \frac{1}{n} \sum_{i=1}^{n} \left| y_i - \hat{y}_i \right|
\end{equation}
where $y_i$ is the true value, $\hat{y}_i$ is the predicted value, and $n$ is the total number of predictions.

\subsubsection{Mean Squared Error (MSE)}
Mean Squared Error measures the average of the squared differences between the predicted and actual values. It is calculated as:
\begin{equation}
\text{MSE} = \frac{1}{n} \sum_{i=1}^{n} (y_i - \hat{y}_i)^2
\end{equation}
where $y_i$ is the true value, $\hat{y}_i$ is the predicted value, and $n$ is the total number of predictions.

\section{Results}\label{result}

Table~\ref{tab:full_results} presents the prediction results of various models in terms of MAE and MSE, evaluated across different input and output sliding window sizes. For each scenario, the model with the lowest error is highlighted in bold.

\vspace{40pt}

\begin{longtable}{|c|c|c|c|c|}
\caption{Performance Comparison Across Different Input-Output Window Sizes}
\label{tab:full_results} \\
\hline
\textbf{Input Window} & \textbf{Output Horizon} & \textbf{Model} & \textbf{MAE}  & \textbf{MSE} \\
\hline
\endfirsthead

\hline
\textbf{Input Window} & \textbf{Output Horizon} & \textbf{Model} & \textbf{MAE}  & \textbf{MSE} \\
\hline
\endhead

\hline
\endfoot

\hline
\endlastfoot

\multirow{27}{*}{5 days} 
& \multirow{9}{*}{1 day ahead} 
& Transformer 1 (Encoder Only) & 0.0166 & 0.0004 \\
& & Transformer 2 (Decoder Only) & \textbf{0.0071} & \textbf{0.0001} \\
& & Transformer 3 (Vanilla Transformer) & 0.0092 & 0.0002 \\
& & Transformer 4 (No Embedding) & 0.0141 & 0.0004 \\
& & Transformer 5 (ProbSparse Attention) & 0.0255 & 0.0009 \\
& & LSTM & 0.0359 & 0.0019 \\
& & TCN & 0.0241 & 0.0009 \\
& & SVR & 0.0448 & 0.0028 \\
& & Random Forest & 0.0234 & 0.0009 \\
\cline{2-5}
& \multirow{9}{*}{5 days ahead} 
& Transformer 1 (Encoder Only) & 0.0152 & 0.0004 \\
& & Transformer 2 (Decoder Only) & \textbf{0.0065} & \textbf{0.0001} \\
& & Transformer 3 (Vanilla Transformer) & 0.0169 & 0.0004 \\
& & Transformer 4 (No Embedding) & 0.0171 & 0.0005 \\
& & Transformer 5 (ProbSparse Attention) & 0.0172 & 0.0005 \\
& & LSTM & 0.0439 & 0.0031 \\
& & TCN & 0.0365 & 0.0023 \\
& & SVR & 0.0487 & 0.0035 \\
& & Random Forest & 0.0377 & 0.0025 \\
\cline{2-5}
& \multirow{9}{*}{10 days ahead} 
& Transformer 1 (Encoder Only) & 0.0169 & 0.0006 \\
& & Transformer 2 (Decoder Only) & \textbf{0.0080} & \textbf{0.0001} \\
& & Transformer 3 (Vanilla Transformer) & 0.0217 & 0.0008 \\
& & Transformer 4 (No Embedding) & 0.0197 & 0.0008 \\
& & Transformer 5 (ProbSparse Attention) & 0.0637 & 0.0075 \\
& & LSTM & 0.0541 & 0.0048 \\
& & TCN & 0.0485 & 0.0041 \\
& & SVR & 0.0485 & 0.0035 \\
& & Random Forest & 0.0503 & 0.0044 \\

\hline

\multirow{27}{*}{10 days} 
& \multirow{9}{*}{1 day ahead} 
& Transformer 1 (Encoder Only) & 0.0096 & 0.0002 \\
& & Transformer 2 (Decoder Only) & \textbf{0.0082} & \textbf{0.0001} \\
& & Transformer 3 (Vanilla Transformer) & 0.0087 & 0.0001 \\
& & Transformer 4 (No Embedding) & 0.0084 & 0.0001 \\
& & Transformer 5 (ProbSparse Attention) & 0.0139 & 0.0003 \\
& & LSTM & 0.0297 & 0.0015 \\
& & TCN & 0.0229 & 0.0009 \\
& & SVR & 0.0563 & 0.0042 \\
& & Random Forest & 0.0248 & 0.0010 \\
\cline{2-5}
& \multirow{9}{*}{5 days ahead} 
& Transformer 1 (Encoder Only) & 0.0189 & 0.0006 \\
& & Transformer 2 (Decoder Only) & \textbf{0.0094} & \textbf{0.0001} \\
& & Transformer 3 (Vanilla Transformer) & 0.0149 & 0.0004 \\
& & Transformer 4 (No Embedding) & 0.0167 & 0.0009 \\
& & Transformer 5 (ProbSparse Attention) & 0.0318 & 0.0016 \\
& & LSTM & 0.0418 & 0.0030 \\
& & TCN & 0.0365 & 0.0024 \\
& & SVR & 0.0539 & 0.0041 \\
& & Random Forest & 0.0396 & 0.0027 \\
\cline{2-5}
& \multirow{9}{*}{10 days ahead} 
& Transformer 1 (Encoder Only) & 0.0232 & 0.0010 \\
& & Transformer 2 (Decoder Only) & \textbf{0.0105} & \textbf{0.0002} \\
& & Transformer 3 (Vanilla Transformer) & 0.0279 & 0.0013 \\
& & Transformer 4 (No Embedding) & 0.0241 & 0.0012 \\
& & Transformer 5 (ProbSparse Attention) & 0.1086 & 0.0167 \\
& & LSTM & 0.0539 & 0.0048 \\
& & TCN & 0.0487 & 0.0041 \\
& & SVR & 0.0538 & 0.0041 \\
& & Random Forest & 0.0522 & 0.0048 \\

\hline

\multirow{27}{*}{15 days} 
& \multirow{9}{*}{1 day ahead} 
& Transformer 1 (Encoder Only) & 0.0103 & 0.0002 \\
& & Transformer 2 (Decoder Only) & \textbf{0.0058} & \textbf{0.0001} \\
& & Transformer 3 (Vanilla Transformer) & 0.0101 & 0.0002 \\
& & Transformer 4 (No Embedding) & 0.0125 & 0.0003 \\
& & Transformer 5 (ProbSparse Attention) & 0.0198 & 0.0006 \\
& & LSTM & 0.0279 & 0.0013 \\
& & TCN & 0.0238 & 0.0009 \\
& & SVR & 0.0556 & 0.0041 \\
& & Random Forest & 0.0265 & 0.0012 \\
\cline{2-5}
& \multirow{9}{*}{5 days ahead} 
& Transformer 1 (Encoder Only) & 0.0159 & 0.0005 \\
& & Transformer 2 (Decoder Only) & \textbf{0.0088} & \textbf{0.0001} \\
& & Transformer 3 (Vanilla Transformer) & 0.0215 & 0.0008 \\
& & Transformer 4 (No Embedding) & 0.0177 & 0.0006 \\
& & Transformer 5 (ProbSparse Attention) & 0.0196 & 0.0006 \\
& & LSTM & 0.0499 & 0.0040 \\
& & TCN & 0.0367 & 0.0024 \\
& & SVR & 0.0522 & 0.0040 \\
& & Random Forest & 0.0413 & 0.0030 \\
\cline{2-5}
& \multirow{9}{*}{10 days ahead} 
& Transformer 1 (Encoder Only) & 0.0205 & 0.0009 \\
& & Transformer 2 (Decoder Only) & \textbf{0.0095} & \textbf{0.0002} \\
& & Transformer 3 (Vanilla Transformer) & 0.0243 & 0.0010 \\
& & Transformer 4 (No Embedding) & 0.0229 & 0.0009 \\
& & Transformer 5 (ProbSparse Attention) & 0.0423 & 0.0028 \\
& & LSTM & 0.0543 & 0.0053 \\
& & TCN & 0.0490 & 0.0041 \\
& & SVR & 0.0521 & 0.0039 \\
& & Random Forest & 0.0532 & 0.0050 \\
\end{longtable}

\begin{figure*}[!htbp]
    \centering
    \includegraphics[width=0.8\textwidth]{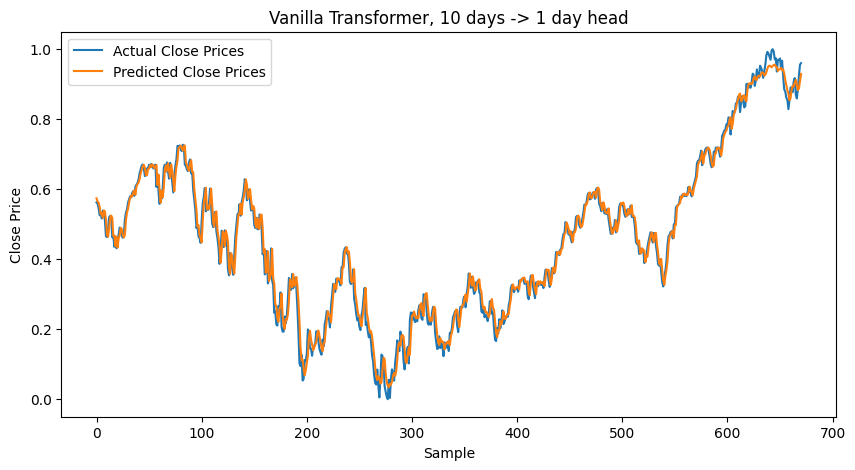}  
    \caption{Transformer Prediction Sample Plot}
    \label{fig: sample plot}
\end{figure*}

 Generally, Transformer models perform better than traditional models like LSTM and TCN. Figure~\ref{fig: sample plot} displays a sample graph of the Vanilla Transformer prediction. As shown, the Vanilla Transformer can accurately capture stock movements. Interestingly, Transformer 2 (Decoder Only) achieves the best performance in all scenarios. This decoder-only design, similar to models like GPT, may be effective at modeling sequences by focusing on past data to predict future values. This approach helps to capture patterns in stock price movements.

Among the transformer variants, Transformer 5 (ProbSparse Attention) performs the worst in most cases. ProbSparse Attention aims to speed up computation by focusing only on the most important parts of the data. However, this can lead to missing some useful information, which may explain its lower accuracy. This trade-off between speed and detail is important to consider when choosing a model.

Another notable finding is that Transformer 4, which lacks embedding layers, outperforms the Vanilla Transformer about half the time. This suggests that, in certain situations, the model can still learn patterns effectively without explicit embeddings. It may be that the model captures temporal information through other means, such as attention mechanisms or the structure of the data itself.

For other deep learning models, TCN generally outperforms LSTM. TCNs use convolutional layers to process sequences, which helps them capture long-term patterns more effectively than LSTMs. This makes TCNs more suitable for time series forecasting tasks.

In traditional machine learning models, Random Forest usually performs better than SVR. Random Forests are ensemble methods that can capture complex patterns and are less prone to overfitting. This makes them more robust for predicting stock prices.

When predicting over a short time horizon, such as one day ahead, traditional machine learning models like Random Forest can outperform deep learning models like LSTM and TCN. However, for longer-term predictions, deep learning models tend to perform better. This is because they can capture more complex patterns over extended periods.

Overall, all models show lower prediction errors when forecasting over shorter time frames. Additionally, transformer models maintain relatively stable performance across different sliding window sizes. This consistency makes them reliable for various forecasting scenarios.

\section{Conclusion}\label{sec5}

This study analyzes the performance of different transformer variants in prediction of S\&P 500 stock index with different sizes of input and output sliding windows. Traditional models such as LSTM, TCN, SVR, and Random Forest are chosen for comparison. The results demonstrate that Transformer structures can outperform other traditional models in time series forecasting. Transformer (Decoder Only) has the best performance in all scenarios. Transformer with ProbSparse attention performs the worst in almost all cases. Transformer without embedding layer can still sometimes outperform standard Transformer model. Transformer models exhibit stable performance in all cases. This study exhibits the superiority of Transformer architectures in time-series forecasting.

\bibliographystyle{unsrt}  
\bibliography{references}

\end{document}